\documentclass{article}

\usepackage{arxiv}

\usepackage[utf8]{inputenc} 
\usepackage[T1]{fontenc}    
\usepackage{hyperref}       
\usepackage{url}            
\usepackage{booktabs}       
\usepackage{amsfonts}       
\usepackage{nicefrac}       
\usepackage{microtype}      
\usepackage{lipsum}
\usepackage{graphicx}
\graphicspath{ {./images/} }

\usepackage{hyperref}
\usepackage{amsmath}
\usepackage{amssymb}
\usepackage{xcolor}
\usepackage{bm}
\usepackage{parskip} 
\usepackage{optidef}
\usepackage{multirow} 
\usepackage{bbold}

\newcommand{\MI}{\text{MI}}

\definecolor{main}{HTML}{5989cf}    
\definecolor{sub}{HTML}{cde4ff}     

\usepackage[many]{tcolorbox}    	

\usepackage{setspace}               
\usepackage{multicol}               

\newtcolorbox{boxC}{
    colback = sub, 
    boxrule = 0pt  
}

\title{On the use of Mutual Information for Testing Independence}

\author{
 Marius Marinescu \\
  Engineering School of Fuenlabrada\\
  King Juan Carlos University \\
  Madrid, Spain \\
  \texttt{marius.marinescu@urjc.es} \\
  \And
  Costel Balcau \\
  Department of Mathematics and Computer Science \\
  Politehnica University of Bucharest \\
  Bucharest, Romania \\
  \texttt{costel.balcau@upit.ro}
}

\begin{document}
\maketitle
\begin{abstract}
    In this paper we use a well know method in statistics, the $\delta$-method, to provide an asymptotic distribution for the Mutual Information, and construct and independence test based on it. Interesting connections are found with the likelihood ratio test and the chi-square goodness of fit test.
    In general, the difference between the Mutual Information evaluated at the true probabilities and at the empirical distribution, can be approximated by the sum of a normal random variable and a linear combination of chi-squares random variables. This summands are not independent, however the normal terms vanishes when testing independence, making the test statistic being asymptotically a linear combination of chi-squares.
    The $\delta$-method gives a general framework for computing the asymptotic distribution of other information based measures. A common difficulty is calculating the first and second-order derivatives, which is already challenging in the case of Mutual Information. However, this difficulty can be circumvallated by using advance symbolic software such as Mathematica.
    Finally, we explore the underlying geometry of the Mutual Information and propose other statical measures which may give competing alternatives to classical tests. 
\end{abstract}

\section{Introduction}



The concept of entropy was introduced by C.E. Shannon in 1948 \cite{shannon1948} within the framework of the mathematical theory of communication. It serves as a measure of the amount of information supplied by a probabilistic experiment or a random variable, drawing inspiration from Boltzmann's entropy \cite{Boltzmann1896} in statistical physics. Building upon Shannon's work, S. Kullback and R.A. Leibler extended this idea by comparing the entropies across a family of probability distributions \cite{kullback1951}, rather than considering the entropy of a single distribution. This led to the introduction of a new measure, known as relative entropy or divergence (KL).
In recent decades, numerous generalizations of entropy have been proposed as information measures. These entropy-based models have found widespread application in various domains of science and technology \cite{leandro2006}.


Mutual Information (MI), a divergence measure, is a fundamental concept in information theory that quantifies the statistical dependence between two random variables. It has been widely employed in various scientific fields, including machine learning, statistical inference, and signal processing, due to its capacity to capture both linear and nonlinear dependencies. Given its importance, the ability to rigorously test independence based on MI is crucial for many applications, particularly in high-dimensional data analysis and complex systems modelling, for instance DNA modelling \cite{menendez2006dna}.

In classical statistics, testing for independence is typically performed using the Pearson’s chi-square test or the likelihood ratio tests, both of which rely on asymptotic distributions derived from categorical data models. However, these tests do not naturally incorporate the information-theoretic perspective offered by MI. A key challenge in utilizing MI for hypothesis testing lies in its estimation from finite samples and the derivation of its asymptotic distribution. Unlike parametric tests, which often assume specific distributional forms, non-parametric MI must account for more general settings, making their theoretical analysis non-trivial.

This paper to fills this gap by leveraging the $\delta$-method to derive an asymptotic distribution for the MI estimator. By applying this method, we establish a theoretical foundation for constructing an independence test based on MI. Specifically, under the null hypothesis of independence, the test statistics asymptotically follows a linear combination of chi-square distributions, which provides a practical means for statistical inference. In addition, our approach reveals interesting connections between MI and the well-known likelihood ratio and chi-square independence tests. 

Furthermore, we explore the MI geometric shape, which sheds light on its behaviour under different probabilistic settings. 
In addition, the methodology employed in this paper, allows for extensions to other information-based measures, thereby offering a broader framework for independence testing. 
Symbolic software may be used for this purpose and for the computation of higher-order derivatives, if necessary.


%

Quantifying and testing dependance is one of the most fundamental tasks in statistics and many measures to study dependencies have been developed in the literature.  $\varphi$-divergences \cite{menendez2006dna, zografos1993} generalises the MI concept by substituting the logarithm by a convex function $\varphi$ which satisfy some other mild conditions. In \cite{leandro1994maestro} a very general measure is presented which encompasses many measures scattered trough the literature, including $\varphi$-divergences, and can be used to perform global studies of most know divergences.
In \cite{hutter2001} it is discussed a Bayesian approach which gives not only a point estimator of MI, but a posterior distribution of it under some prior suppositions.
%
In \cite{ihler2004nonparametric}, a test to determine the dependence structure of a random vector is studied. The problem is formulated as a hypothesis test where each hypothesis represents a different way of grouping variables into independent subsets.  The test can be chosen to be non-parametric and is based on likelihood ratios and KL divergences.
%
In \cite{gabor2007, gabor2009} a measure called distance correlation (dCor) is presented to test dependencies between random vectors. Like MI, the distance correlation is zero if and only if the variables are independent. dCor is based on energy distances \cite{gabor2013energy} and is defined by using the characteristic functions of the joint and marginal distributions (energy distance measured in the Fourier domain). Then, the authors derive an empirical version for practical computation and a test of independence is formulated using a permutation-based approach. Monte Carlo simulations suggests that distance correlation is more powerful than classical tests, when the dependence structure is
nonlinear and similar when is linear.
In \cite{lopez2013} the  Randomized Dependence Coefficient is introduced which uses a empirical copula to measure non-linear dependencies between random vectors of arbitrary dimension with a low complexity $O(n\log n)$.
In \cite{reshef2011}, the maximal information coefficient is presented. Some time after, it has been reported to have less power \cite{simon2014comment,gorfine2012comment} in detecting some associations, including linear.
Finally, in \cite{arthur2005} an independence test based on the Hilbert-Schmidt Independence Criterion (HSIC) is proposed. HSIC involves constructing an appropriate kernel matrix and the use of the Frobenius Norm to summarize the dependency of the data.
In general, there is no single best independence test for all cases, and very complex tests may loose interpretability \cite{reimherr2013}.

The remainder of the paper is organized as follows. Section \ref{sec:1} presents the derivation of the MI asymptotic distribution using the $\delta$-method. Section \ref{sec: continuous} extends the results to the continuous case. In Section \ref{sec:discussion}, we discuss the implications of our findings, compare MI-based tests with classical independence tests, and highlight potential areas for further research. Finally, Section \ref{sec:conclusion} concludes the paper.

\section{Derivation of the asymptotic distribution}\label{sec:1}

In this section, a asymptotic distribution for the MI is derived, based on the $\delta$-method. The discrete (finite) case is treated in this section. Then, in Section \ref{sec: continuous} a natural generalisation to the continuos case is presented.

%

Consider we have an i.i.d. sample of $n$ observation from a pair of random variables $(X,Y)$. Suppose that the support of the random variable is $\Omega_{X,Y}=\{(x_1,y_1), (x_2,y_1), \dots, (x_{I},y_1),(x_1,y_2), (x_2,y_2), \dots, (x_I,y_J) \}$. We will use the following short-hand notation:  $p_{ij}=P(X=x_i, Y=y_j)$, $p_{i*}=\sum_{j=1}^J p_{i j}$ and $p_{* j}=\sum_{i=1}^I p_{i j}$, for $i =1,\dots ,I$, and $j =1,\dots,J$.  We can represent the distribution in a rectangular table as shown in Table \ref{tab:contingency_table}, which is usually known as contingency table. 
\begin{table}[ht]
\centering
\[
\begin{array}{|c|c|c|c|c||c|}
\hline
P_{X,Y} & j=1 & j=2 & \cdots & j=J  & \text{Marginal} \ P_X \\ \hline
i=1 & p_{11} & p_{12} & \cdots & p_{1J} & p_{1*}\\ \hline
i=2 & p_{21} & p_{22} & \cdots & p_{2J} & p_{2*} \\ \hline
\vdots & \vdots & \vdots & \ddots & \vdots & \vdots\\ \hline
i=I & p_{I1} & p_{I2} & \cdots & p_{IJ} & p_{I*} \\ \hline \hline
\text{Marginal} \ P_Y & p_{*1} & p_{*21}  & \cdots & p_{*J} & p_{**}=1 \\ \hline
\end{array}
\]
\caption{Two-way contingency table for a bivariate mass function \( P_{X,Y} \).}
\label{tab:contingency_table}
\end{table}
Since we only have a sample, and we don't know any $p_{ij}$ we have to work with relative frequencies, also known as the empirical distribution or empirical processes \cite[Chapter 19]{vaart2000}, \cite[Section 3.8]{vaart1996}. Let $n_{ij}$ denote the observed frequency of the value $(x_i,y_j)$. Then, the non-parametric estimator of the probability $p_{ij}$ is: $\hat{p}_{ij}=\frac{n_{ij}}{n}$. Let $\bm{p}$ and $\hat{\bm{p}}$ design the column vectorisation of the true distribution and the relative frequency, respectively.
Finally, we denote with $\bm{p}^*$ the column vectorisation of the distribution constructed as the product of the marginal and $\hat{\bm{p}}^*$ the column vectorisation of his likelihood estimator through the sample. 
We want to test the hypothesis that the pair of random variables $X,Y$ are independent, which gives the null hypothesis:
\begin{equation}
H_0: p_{i j}=p_{i *} p_{* j}, \quad i=1, \ldots, I, \quad j=1, \ldots, J
\end{equation}
against the negation of it. To test this hypothesis, we take advantage of the property that MI is zero if and only if the random variables are independent. Given an asymptotic distribution for the MI estimator, we could directly construct the hypothesis test based on it. Let's see how we can construct this asymptotic distribution. The MI is defined as:
\begin{equation}\label{eq: MI}
    \MI(\bm{p})= \sum_{i,j=1}^{I,J} p_{ij}\ln \frac{p_{ij}}{p_{i*}p_{*j}}. 
\end{equation}
The MI can be seen as a Kullback–Leibler divergence measure between the joint probability distribution ($\bm{p}$) and the ones constructed as the (outer) product of the marginals ($\bm{p}^*$) \cite{kullback1997}. 

To find an asymptotic distribution for $\MI(\hat{\bm{p}}_n)$, the ``plug-in'' or substitution estimator of MI, we will use the $\delta$-method \cite{ver2012, oehlert1992}, \cite[Section 6a.2]{rao1973}. The key idea of the $\delta$-method is to use a Taylor expansion, which for the case of MI centred at $\MI(\bm{p})$ can be written as:
\begin{align}\label{eq: Tn_Taylor}
    \MI(\hat{\bm{p}}_n)= \MI(\bm{p}) & + \nabla \MI \bigg|_{\bm{p}} (\hat{\bm{p}}_n- \bm{p}) \\
    & +\frac{1}{2} (\hat{\bm{p}}_n- \bm{p})^\top H \bigg|_{\bm{p}} (\hat{\bm{p}}_n- \bm{p}) +  R(|| \hat{\bm{p}}_n-   \bm{p}||^2) \notag
\end{align}

where $\nabla \MI =\{ \frac{\partial \MI}{\partial p_{ij}} \}_{\substack{i=1,...,I;\\ j=1,...,J}}$ is the gradient vector, $H= \{\frac{\partial\MI}{\partial p_{st}p_{ij}} \}_{\substack{s,i=1,...,I;\\ t,j=1,...,J}}$ is the Hessian matrix and $R$ is a rest function such that $\frac{ R(|| \hat{\bm{p}}_n-  \bm{p}||^2 )}{|| \hat{\bm{p}}_n-   \bm{p}||^2} \xrightarrow[n \to \infty]{P} 0$. This expansion applies for $C^2$ functions. MI is $C^\infty$ on the domain $0<p_{ij} \le 1$.


%

The $\delta$-method requires that $(\hat{\bm{p}}_n- \bm{p} )\xrightarrow[n \to \infty]{L} \text{Multivariate Normal}$. This is actually the case for the empirical distribution whose probability law is a scaled binomial, and converge to a normal by the central limit theorem. See for instance \cite[Section 14.1.4]{agresti2012}, for more information. In general, we have the following result which is going to be used:
\begin{equation}\label{eq: normal}
    \sqrt{n}(\hat{\bm{p}}_n- \bm{p} )\xrightarrow[n \to \infty]{L} N(\bm{0}, \Sigma)
\end{equation}
\noindent with $\Sigma=\text{diag}(\bm{p} ) - \bm{p} \bm{p}^\top$.

The $\delta$-method is typically described as a first order expansion. We are going to see that for the case of testing independence with the MI, the gradient vanishes thus being necessary a second or superior order of approximation to find an (non-degenerate) asymptotic distribution. See \cite[Section 20.1.1 and Theorem 20.8]{vaart2000} for more information about higher order $\delta$-methods.

Since, the distribution of $\sqrt{n}(\hat{\bm{p}}_n- \bm{p})$ is asymptotically normal it results that if in the Taylor expansion, given in eq. \ref{eq: Tn_Taylor}, we move $\MI(\bm{p})$ to the left side and multiply both sides by $\sqrt{n}$ we obtain
on the right a linear combination of asymptomatic normal variables plus a quadratic of the same asymptomatic normals (and a rest that tends to zero in probability). Since the linear term vanishes in our case study, a quadratic form remains. A quadratic form of a normal random vector is always a linear combination of Chi random variables, and for some specific cases is exactly a Chi distribution (see \cite[Chapter 4]{mathai1992} or
\cite[Section 20.5.2]{seber2008matrix}). 

We have to be cautious about using the MI expression as presented in eq. \ref{eq: MI} or in the literature. Actually the Mutual Information is not a function of $I \cdot J$ variables but of $I \cdot J-1$ since we have the restriction $\sum_{i,j=1}^{I,J} p_{ij} = 1$. When the random variables are independent we have the additional restrictions $p_{ij}=p_{i*}p_{*j}$ for all $ i =1,\dots ,I$ and $j =1,\dots,J$. Since $\MI(\hat{\bm{p}}_n)$ is a function of the sample distribution, the last restriction is not necessarily satisfied even when the random variables are independent, whereas the first normalisation restriction always applies. Thus, when computing $\MI(\hat{\bm{p}}_n)$ derivatives, we will substitute $p_{IJ}$ by $1 - \displaystyle\sum_{\substack{i,j=1, \  i \neq I, \ j \neq J}}^{I,J} p_{ij} $.


After same careful calculations, it can be seen that the partial derivatives of $\MI(\hat{\bm{p}}_n)$  are (see Appendix~\ref{ap:grad}):
\begin{equation}
    \frac{\partial \text{MI}}{\partial p_{ij}}= \ln \frac{p_{ij}}{p_{i*}p_{*j}} - \ln \frac{p_{IJ}}{p_{*J}p_{I*}},  \ \forall i=1,...,I, \ \forall j=1,...,J, \text{ and } (i,j) \neq (I,J).
\end{equation}

\noindent Clearly, under the hypothesis $H_0$ the previous partial derivatives vanishes, making $\nabla \MI =\bm{0}$. Thus, under $H_0$  eq. \ref{eq: Tn_Taylor} simplifies to:

\begin{equation}
    \MI(\hat{\bm{p}})= \MI(\bm{p}) + \frac{1}{2} (\hat{\bm{p}}_n- \bm{p})^\top H \bigg|_{\bm{p}} (\hat{\bm{p}}_n- \bm{p}) +  R(|| \hat{\bm{p}}_n-   \bm{p}||^2).
\end{equation}

Now, the Hessian matrix remains to be computed. 
The computation of the Hessian matrix is more challenging and requires some patience and attention to detail. In Appendix \ref{ap:hess} the computation is given. The general expression of H dependents on four indexes and can be described as:

\begin{itemize}
    \item $s=i, \ t=j$, \ $\frac{\partial^2\text{MI}}{\partial p_{st}\partial p_{ij}} + a(s,t) = \begin{cases}
        \frac{1}{p_{ij}}-\frac{1}{p_{i*}}-\frac{1}{p_{*j}} & \ \ i \neq I, \ j\neq J\\
         \frac{1}{p_{ij}}-\frac{1}{p_{*j}} & \ \ i = I, \ j\neq J\\
          \frac{1}{p_{ij}}-\frac{1}{p_{i*}} & \ \ i \neq I, \ j= J\\
    \end{cases}$
    
    \item $s=i, \ t \neq j$, \  $\frac{\partial^2\text{MI}}{\partial p_{st}\partial p_{ij}} + a(s,t)= \begin{cases}
        \frac{1}{p_{i*}} & \ \ i \neq I, \ j\neq J\\
         0 &  \ \ i = I, \ j\neq J\\
          -\frac{1}{p_{i*}} + \frac{1}{p_{*j}} & \ \ i \neq I, \ j= J\\
    \end{cases}$
    
    \item $s \neq i,\  t = j$, \  $\frac{\partial^2\text{MI}}{\partial p_{st}\partial p_{ij}} + a(s,t)= \begin{cases}
        \frac{1}{p_{*j}} & \ \ i \neq I, \ j\neq J\\
          \frac{1}{p_{i*}} - \frac{1}{p_{*j}} & \ \ i = I, \ j \neq J\\
          0 & \ \ i \neq I, \ j = J\\
          \end{cases}$

    \item $s \neq i,\  t \neq j$, \ 
 $\frac{\partial^2\text{MI}}{\partial p_{st}\partial p_{ij}} + a(s,t)= \begin{cases}
       0 & \ \ i \neq I, \ j\neq J\\
          \frac{1}{p_{i*}} - \frac{1}{p_{*j}} & \ \ i \neq I, \ j= J\\
          0 & \ \ i = I, \ j\neq J\\
    \end{cases}$
\end{itemize}

with 
\begin{equation*}
    a(s,t)=\begin{cases}
        \frac{1}{p_{IJ}}-\frac{1}{p_{I*}}-\frac{1}{p_{*J}} & \ \ s \neq I, \ t\neq J\\
         \frac{1}{p_{IJ}}-\frac{1}{p_{*J}} & \ \ s = I, \ t\neq J\\
          \frac{1}{p_{IJ}}-\frac{1}{p_{I*}} & \ \ s \neq I, \ t= J \ .\\
    \end{cases}
\end{equation*}

The Hessian matrix can be written in a somewhat more compact form as:

\begin{equation}\label{eq: hess}
    H = A_{(I-1)J \times IJ}    \cdot \begin{pmatrix}
         1 &  0 & \cdots & 0 \\
         0 & 1 &  \cdots & 0\\
         \vdots & \vdots & \ddots & \vdots \\
        0 & 0 & \cdots  & 1 \\
        -1 & -1 & \cdots & -1 \\
    \end{pmatrix}_{IJ \times (I-1)J} - \begin{pmatrix}
        1 \\
        1 \\
        \vdots \\
        1
    \end{pmatrix}_{(I-1)J \times 1} \text{vec}_2(\begin{pmatrix}
        B & \bm{c}_2 \\
        \bm{c}_1^\top & 0
    \end{pmatrix})^\top
\end{equation}

with 
\begin{equation*}
     A_{ij, st}=\begin{cases}
        \frac{1}{p_{ij}}-\frac{1}{p_{i*}} -\frac{1}{p_{*j}} & \ \ s = i,\ t=j \\
          -\frac{1}{p_{i*}} & \ \ s = i,\ t \neq j  \\
           -\frac{1}{p_{*j}}  & \ \ s \neq i,\ t=j  \\
          \end{cases}
\end{equation*}
\begin{equation*}
    B= ( \frac{1}{p_{IJ}}-\frac{1}{p_{I*}} -\frac{1}{p_{*J}}) \cdot \mathbb{1}
\end{equation*}
\begin{equation*}
    \bm{c}_1=  ( \frac{1}{p_{IJ}} -\frac{1}{p_{*J}}) \cdot \bm{1}
\end{equation*}
\begin{equation*}
    \bm{c}_2= ( \frac{1}{p_{IJ}}-\frac{1}{p_{I*}}) \cdot \bm{1}
\end{equation*}
and vec$_2$ stands for the column vectorization operator with the last element removed.

As we see, the Hessian matrix depends on multiplicative inverses of the joint and marginal pmf's values.
We have supposed that all $p_{ij}>0$, which is always the case when the random variables are independent. In practice, only the values which have been observed ($\hat{p}_{ij}>0$) are used for the test.

\begin{figure}[ht]
    \centering
    \includegraphics[width=0.52\linewidth]{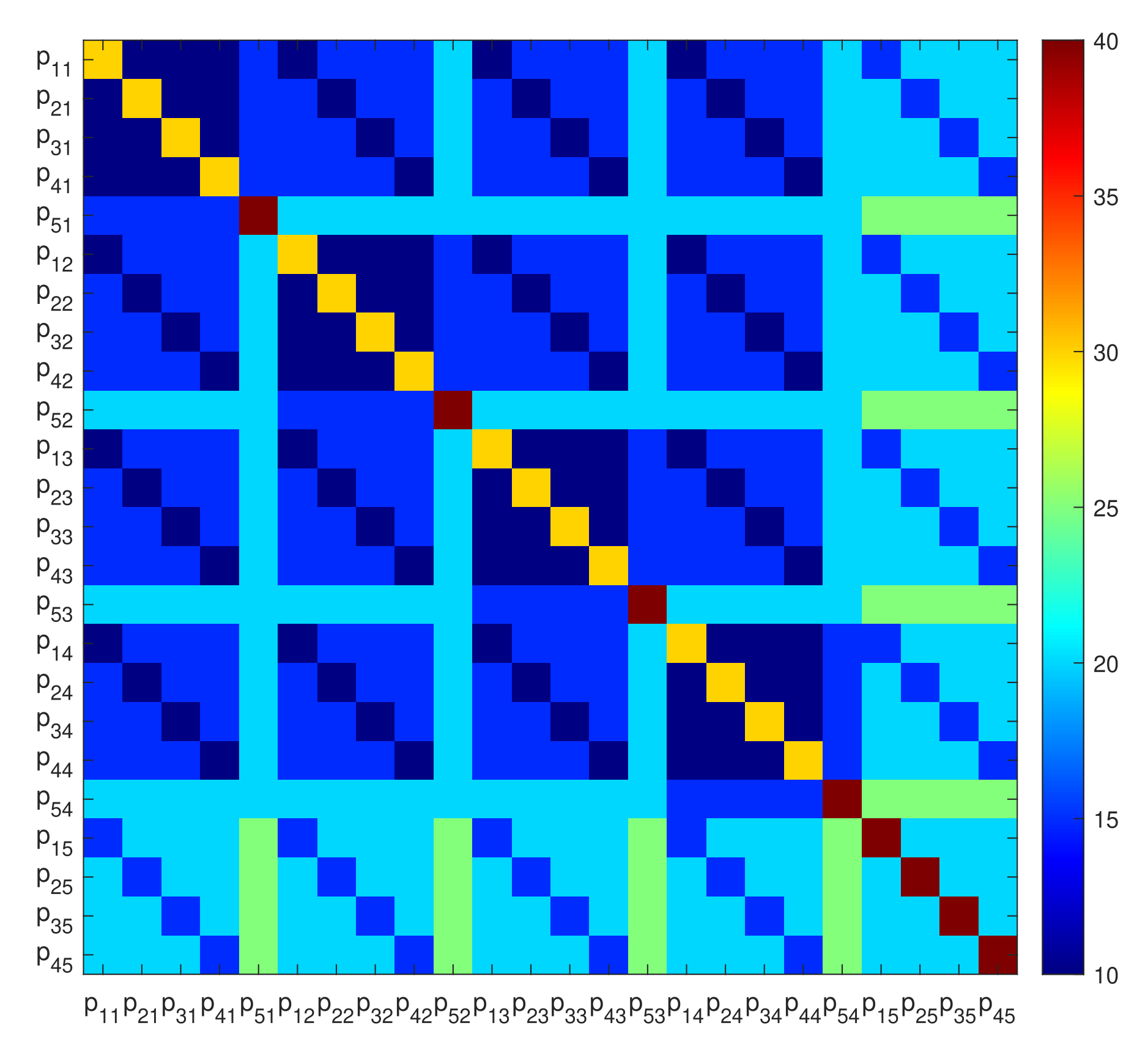}
    \caption{The Hessian matrix of the MI of two independent Uniform random variables in $\{0,1,2,3,4\}$. }
    \label{fig:hess_mat}
\end{figure}

In Fig. \ref{fig:hess_mat} the Hessian matrix of a bivariate uniform random variable is shown. It can be seen that the matrix is symmetric and have interesting visual patterns.
Now that we have the expression of all the elements in eq. \ref{eq: Tn_Taylor} we may construct an hypothesis test. First, we multiply both sides by $2n$ and rewrite the equation as:
\begin{equation}\label{eq:MI2}
    2n(\MI(\hat{\bm{p}}_n) -\MI(\bm{p})) =  n(\hat{\bm{p}}_n- \bm{p})^\top H \bigg|_{\bm{p}} (\hat{\bm{p}}_n- \bm{p}) +  R(|| \hat{\bm{p}}_n-  \bm{p}||^2).
\end{equation}

Taking into account eq. \ref{eq: normal}, the right-hand side of eq. \ref{eq:MI2} is a quadratic form in a asymptotic multivariate normal variables plus a rest that tends to zero in probability. By \cite[Theorem 
 20.28]{seber2008matrix} the quadratic form is asymptotically distributed as:
\begin{equation}
     \sqrt{n}(\hat{\bm{p}}_n- \bm{p})^\top H \bigg|_{\bm{p}} \sqrt{n}(\hat{\bm{p}}_n- \bm{p})  \xrightarrow[n \to \infty]{L} \bm{\lambda}^\top \begin{pmatrix}
         \bm{\chi^2}
     \end{pmatrix}
\end{equation}
with $\bm{\lambda} = \text{tr}(\Sigma H)=\text{tr}(H \Sigma)$ and $\bm{\chi^2}$ is a column vector of $I \cdot J-1$ i.i.d. chi-square random variables with one degree of freedom. That is, the quadratic form is asymptotically distributed as a linear combination of chi-square random variables. This distribution is not very common in the literature but a way to calculate its cdf to a given degree of accuracy, by using an infinite gamma series, can be found in \cite{moschopoulos1984}. Alternatively, the quantiles of interest can be easily generated by simulation. Let's denote by $\chi_{\bm{\lambda}}^2$ this distribution.



Thus, we have finally found a asymptotic distribution to perform the hypothesis test. Under the null hypothesis $\bm{p} := \bm{p}_{H_0} = \text{vec}_2(\bm{p}_X \cdot \bm{p}_Y^\top)$, where $\bm{p}_X$, $\bm{p}_Y$ represent the theoretic marginal distribution of $X$ and $Y$, respectively. To perform the hypothesis test two candidates arises:
\begin{align}
    T^1_n &= 2n(\MI(\hat{\bm{p}}) - \MI(\bm{p}_{H_0}))   \stackrel{H_0}{=}
 2n\MI(\hat{\bm{p}})  \text{ and } \\
    T^2_n &= n(\hat{\bm{p}}_n-  \bm{p}_{H_0})^\top H \bigg|_{\bm{p}_{H_0}} (\hat{\bm{p}}_n-  \bm{p}_{H_0}).
\end{align}
The difference between them is of order $o_p(|| \hat{\bm{p}}_n - \bm{p}||^2)$ and tends to zero in probability when $n$ grows.

\begin{figure}[ht]
    \centering
    \includegraphics[width=1\linewidth]{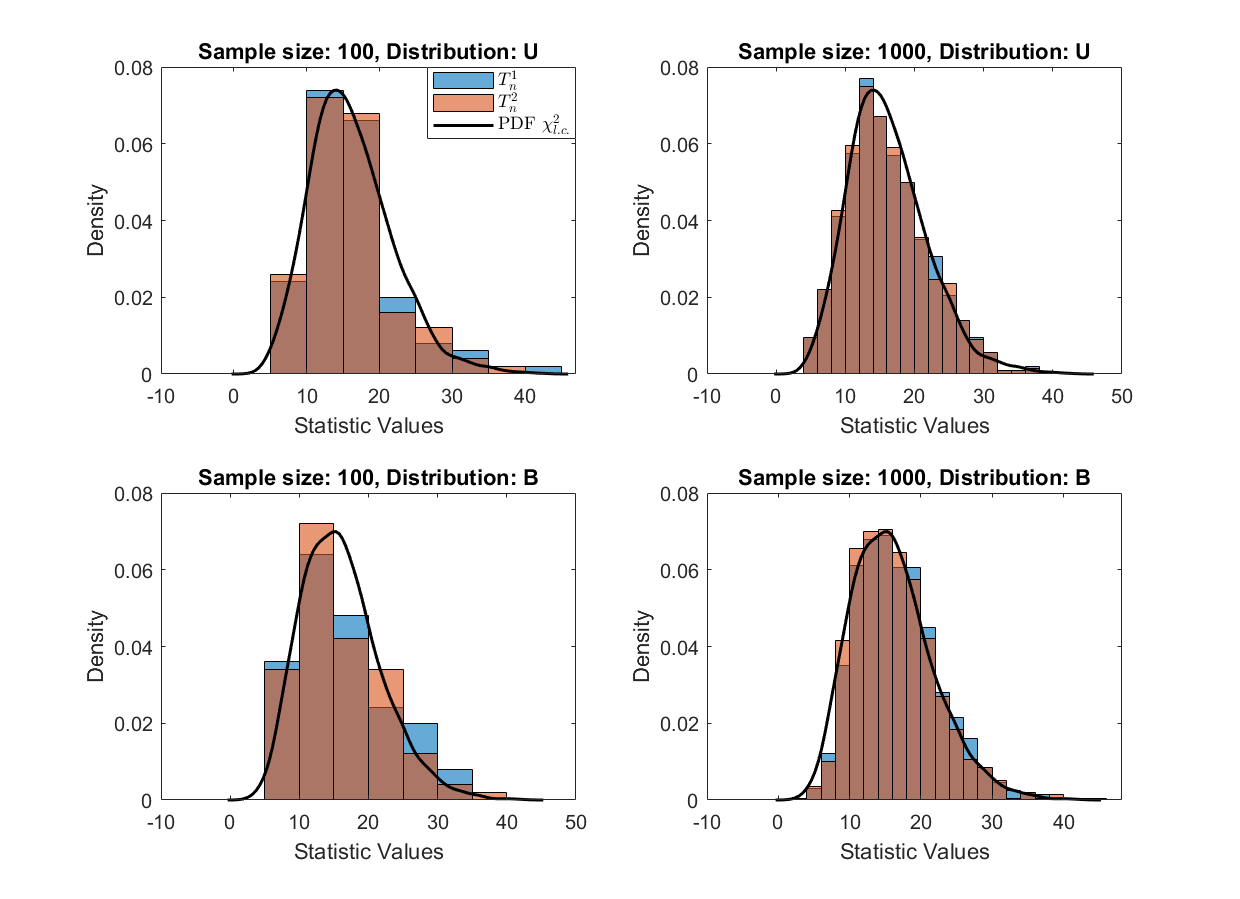}
    \caption{Histogram of $T^1_n$ and $T^1_n$ for sample sizes 100 (left) and 1000 (right) for the bivariate independent Uniform and Binomial distribution. U stands for $U\{0,1,2,3,4\}$ and B for $B(4,0.5)$. The pdf of $\chi_{\bm{\lambda}}^2$ have been retrieved by Kernel density estimation.}
    \label{fig:Tsim}
\end{figure}

In Figure \ref{fig:Tsim} the histograms of both statistics have been computed by simulation. In particular, the Uniforms and Binomial distributions have been sampled. It can be observed that the distribution of the statistics converges when sample size increases and that the pdf $\chi_{\bm{\lambda}}^2$ is a good fit, as expected from the previous theoretic derivations.

Is interesting to notice that $T_n^1$ does not dependent on the unknown values $\bm{p}_{H_0}$, nevertheless $T_n^2$ does. 
%
%
Thus, for $T_n^1$ we may construct an hypothesis test by considering the acceptance region $ R_0 = \{ T_n^1 \in \mathbb{R}^+: T_n^1 < \chi_{\bm{\lambda}, \alpha} \}$ with $P (\chi_{\bm{\lambda}} < \chi_{\bm{\lambda}, \alpha})=1-\alpha$.
For $T_n^2$ we have that the statistic depends on the true marginal distribution $\bm{p}_{H_0}$ which is typically unknown. A practical solution is to consider a plug-in estimator and substitute  $\bm{p}_{H_0}$ by the product of the sample marginals, $\hat{\bm{p}}_{H_0}$. In that case, the substituted term is no longer a constant know value and convergence should be revisited. There exists generalisations of the $\delta$-method that allow to substitute $\bm{p}_{H_0}$ by a statistic who converge to $\bm{p}_{H_0}$. A sufficient condition is that the derivatives of MI are uniform differentiable \cite[Theorem 3.9.5]{vaart1996}. Since MI has continuos derivatives in a neighbourhood of $\bm{p}_{H_0}$, it is uniformly differentiable. See also, \cite[Note on pg. 386 and eq. 6a.2.5]{rao1973} for generalisations of the delta method to non-constant but convergent parameters and \cite[Note 14.2. on pg. 594]{agresti2012} for generalisation of the $\delta$-method to higher-orders when first derivatives vanishes.
%
%
Notice that marginals convergence, $\hat{\bm{p}}_{H_0} \to \bm{p}_{H_0}$, is faster than joint convergence $\hat{\bm{p}} \to \bm{p}$ \cite{agresti2012}.
%
%
%
Note also that $T_n^2$ uses more information than $T_n^1$ in the sense that it uses both the joint distribution and the marginal one in his definition.



Finally, notice that we have used three approximation to derive the test: 1) the effect of $R(|| \hat{\bm{p}}_n-  \bm{p}_{H_0}||^2)$ have been neglected, 2) considering normality of $\sqrt{n}(\hat{\bm{p}}_n- \bm{p})$ which is only asymptotic, and 3) for $T_n^2$ approximating $\bm{p}_{H_0}$ by $\hat{\bm{p}}_{H_0}$.


\subsection{Connections with other independence test}

Detecting statistical dependence have been widely treated in the literature. One of the primary works on the topic relays on the work of Pearson in 1900, who described the chi-square goodness of fit test based on a geometric approach \cite{pearson1900}. He defined a frequency surface for the deviation from the mean, and described $\chi^2=\text{cte.}$ as a generalised ellipsoid \cite[eq. II]{pearson1900}.
Two decades after (with years of controversy \cite{baird1983}), Fisher \cite{fisher1922} corrected the chi-square test by reducing the degrees of freedom (term coined by him) from $IJ-1$ to $(I-1)(J-1)$ for the case of testing independence. This corrections was specially important for examples with reduced number of rows and columns in the contingency tables. By overestimating the degrees of freedom the chi-square critical values were larger and p-values inflating.
%
Another, know statistic is the likelihood ratio test presented in 1933 by Neyman and Pearson's son \cite{neyman1933, cressie1989}.
This two are the best know statistics for testing independence and their presentation is standard in categorical data analysis books.

The likelihood ratio test is based on the ratio \cite[Section 3.2.]{agresti2012}:
\begin{equation}
\Lambda=\frac{\prod_i \prod_j\left(n_{i*} n_{*j}\right)^{n_{i j}}}{n^n \prod_i \prod_j n_{i j}^{n_{i j}}}_. 
\end{equation}

This computes the ratio between the multinomial likelihoods of the hypothesised counts for each cell under $H_0$ and the multinomial likelihood of the actual counts. It is know that the asymptotic distribution of $G^2=-2\ln{\Lambda}$ is $\chi_{(I-1)(J-1)}$ \cite[Section 14.3.4.]{agresti2012}. Rearranging, we get the typically description of $G^2$,
\begin{equation}
G^2=-2 \ln \Lambda=2 \sum_i \sum_j n_{i j} \log \left(n_{i j} / ((n_{i*} n_{*j})/n) \right).
\end{equation}
%
Dividing by $n$ on the numerator and denominator inside the logarithm, and dividing and multiplying by $n$ outside we get the expression of $G^2$ in terms of the empirical distribution

\begin{equation}
G^2= 2n\sum_i \sum_j \hat{p}_{i j} \ln \left(\hat{p}_{i j} / (\hat{p}_{i*} \hat{p}_{*j}) \right) = 2n\MI(\hat{\bm{p}}) = T_n^1.
\end{equation}

which is exactly the statistics $T_n^1$ derived from the MI. Thus, performing the test with $T_n^1$ is equivalent to performing an independence log-likelihood ratio test.

On the other hands, a very interesting connection appears between $T_n^2$ and the classical Chi-square test for independence,
\begin{equation}
    \chi^2=\sum_{ij=1}^{IJ} \frac{\left(n\hat{p}_{ij}-n \hat{p}_{i*}\hat{p}_{*j}\right)^2}{n \hat{p}_{i*}\hat{p}_{*j}}_.
\end{equation}

I have the conjecture that, under $H_0$ and the restriction $\sum_{i,j=1}^{I,J} p_{ij}=1$, the expression of $T_{n}^2 $ reduce to $ \chi^2$.
The computation to show that $T_n^2 \stackrel{H_0}{=} \chi^2$ are not directly tractable by hand (a quadratic in $IJ-1$ variables) and other methods may apply which remains an open work\footnote{This connection have been verified numerically and symbolically and is presented as a supplementary material of this paper.}. In \cite[Eq. 1.16]{agresti2012} a similar problem is described with $H$ changed by $\Sigma_*^{-1}$ with $\Sigma_*$ being the covariance matrix $\Sigma$ with the last row and column removed which makes it full rank.

%

Thus, it results that MI, which has the property that $\MI=0$ $\iff$ $X$ and $Y$ are independent, connects in an subtle and elegant way the two best known state of the art independence test statistics, which evidence the importance of the MI concept.

\section{Continuos case}\label{sec: continuous}

For the continuos case we have:

\begin{equation}
\MI(f)=\int \int_{\Omega_{X,Y}} f_{X, Y}(x, y) \log \left(\frac{f_{X, Y}(x, y)}{f_X(x) f_Y(y)}\right) d x d y.
\end{equation}

$\MI(f)$ and $\MI(\bm{p})$ are equivalent definitions in the sense that there exist a limiting processes that relates both\footnote{Note this is not the case of the entropy, where the discrete version and the continuos counterpart called differential entropy are not equivalent and have different properties \cite[Section 9.3]{thomas2006}.} \cite[Section 9.5]{thomas2006}. Thus, a natural approach is to divide the support of the continuos bivariate random variable $(X,Y)$ in a partition, compute the relative frequency in each bin and apply the discrete test as described in Section \ref{sec:1}. A way to chose an adequate partition size is the Pearson square root or the Rice rule \cite{rubia2024hist}, among others.

An alternative way is to estimate the joint density $\hat{f}$, for instance, by a kernel density method. Then choose a partition $R_{ij}=\{A_i \times B_j\}_{\substack{i=1,...,I;\\ j=1,...,J}}$ and evaluate $\hat{p}_{ij} :=\int \int_{R_{ij}} \hat{f} \ dA= \hat{f}(x^*_i,y^*_j) A_{ij}$ for some $(x^*_i, y^*_j) \in R_{ij}$ and where $A_{ij}$ represent the area of the $(i,j)$ element of the partition. The last equality follows by the mean value Theorem, whenever $f$ is continuos. Then apply the $\delta$-method to
\begin{equation}
    \MI(\hat{f}) = \sum_{i,j=1}^{I,J} \int \int_{R_{ij}} \hat{f} \ dA =  \sum_{i,j=1}^{I,J} \hat{p}_{ij}\ln \frac{\hat{p}_{ij}}{\hat{p}_{i*}\hat{p}_{*j}} 
\end{equation}

where the last expression comes from multiplying and dividing by $A_{ij}$ inside the logarithm. 









 


\section{Discussion}\label{sec:discussion}

MI simplifies the independence property in such a way that testing independence is equivalent to test if a statistic is equal to zero vs the alternative hypothesis it is greater. By using the $\delta$-method an asymptotic distribution have been derived making it possible to test independence. Two statistics, $T_n^1$ and $T_n^2$, naturally arises which result to be equivalent to known statistics in the categorical data analysis field. 

In practice $T_n^2$ typically is preferred to $T_n^1$.
For a fix number of bins, as $n$ becomes larger, the distribution of $T_n^2$ typically approaches a chi-squared distribution faster than that of $T_n^1$. The chi-squared approximation is often poor for $T_n^1$ when $n/(IJ) < 5$. When $IJ$ is large, this approximation can be relatively good for $T_n^2$ even if $n/(IJ) \le 1$, provided the table does not simultaneously have very small and moderately large expected frequencies.
$T_n^2$ applies with smaller $n$ and more sparse tables than $T_n^1$. Depending on the sparseness of the table $T_n^1$ has higher risk that p-values based on referring $T_n^1$ to a chi-squared distribution to be too large or too small. See \cite{cressie1989, mchugh2013chi} for more details.



%

\begin{figure}[ht!]
    \centering
    \includegraphics[width=0.75\linewidth]{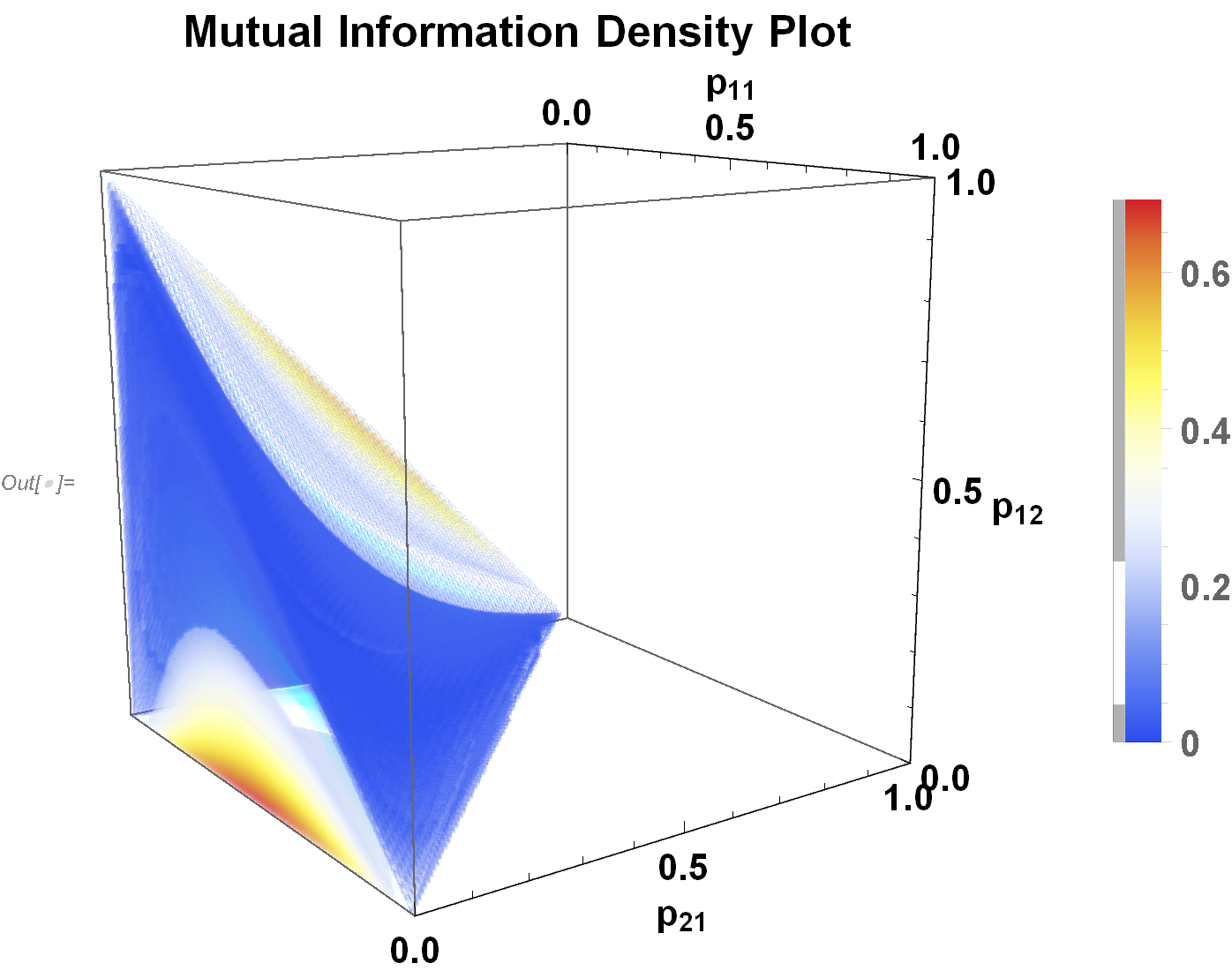}
    \caption{MI as a function of the probabilities $p_{11}, p_{21}, p_{12}$ in a $2\times 2$ contingency table. $p_{22}$ is excluded since $p_{22}=1 - p_{11} - p_{21} - p_{12}$.}
    \label{fig:MI22}
\end{figure}

\begin{figure}[h]
     \centering
\includegraphics[width=0.6\linewidth]{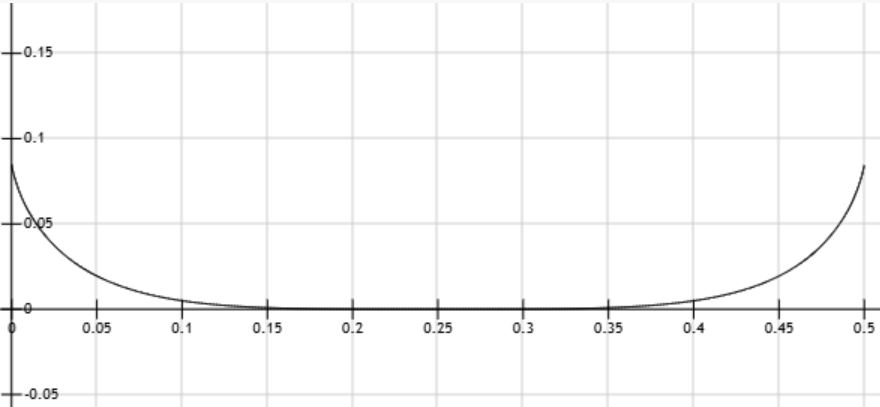}
     \caption{MI as a function of $p_{11}$, for the discrete uniform $ 2\times 2$ bivariate random variable. }\label{fig:MI2x2}
 \end{figure}

MI does not inherently take into account ordinal or interval relationship within categories. This is because MI is based solely on the joint and marginal probability distributions of the variables, without any explicit consideration of the order or spacing between categories. For instance, a bivariate random variable on a circular support is inherently depended, nevertheless, MI does not detect this.
MI is a divergence based measure, that it, MI measures the ``distance'' between the joint distribution and the one composed by the product of the marginals, whatever the support is. For the sake of illustration, in Fig. \ref{fig:MI22} this divergence is represented for a general $2\times 2$ contingency table. The shape of the MI function can be observed. In blue we can identify the level surface MI=0. The acceptance region $R_0$ can be imagined as a confidence region around it. 
In some cases, MI can be very conservative for testing independence. For example, consider a uniform discrete $ 2\times 2$ bivariate random variable except we let $p_{11}$ free to move. The MI as a function of $p_{11}$ is represented in Fig. \ref{fig:MI2x2}. The possible range for $p_{11}$ to be a valid distribution is [0, 0.5]. We may observe that MI is almost tangent to $y=0$ for values $0.1 \le p_{11} \le 0.4$. This makes harder for the test to rejected independence for values $p_{11} \in [0.1, 0.4]$, despite the only value that makes the distribution independent is $p_{11}=0.25$.
Other divergence-based measure have other geometries which may change the power and (true) significance level of the test to be used.
Using the $\delta$ method, other statistics can be considered to test independence and symbolic computation can be used to compute any gradient and Hessian matrix for more complex measures. For instance, the normalised mutual information \cite{studholme1999overlap} can be used  $\text{NMI}:= \frac{\MI(\bm{p})}{H(p_{X,Y})}$, where $H(p_{X,Y})$ is the joint entropy. 
    NMI takes into account the proportion of shared information relative to the total information contained in the joint distribution. In addition, it is a bounded distance and is considered more robust in the sense that handles cases where the variables have vastly different entropies or scales.
    Generalisation of the NMI \cite[Eq. 16]{marius2011} can be considered using Tsallis entropy \cite{tsallis1988}, which comes from non-extensive statistical mechanics. Robust results for measuring scanned documents similarity applied to classification have been reported in \cite{marius2011} for Tsallis entropy.

\vspace{1cm}




\section{Conclusion}\label{sec:conclusion}

In this work, we derived the asymptotic distribution of the MI sample estimator under the hypothesis of independence, using the $\delta$-method and established its connection to classical independence tests. 
In particular, we have shown that under the independence hypothesis, $2n\MI$ is asymptotically a linear combination of chi-squares random variables and that MI gives rise to both the log-likelihood ratio $(T_n^1)$ and the chi-square $(T_n^2)$ statistics. The difference between $T_n^1$ and $T_n^2$ is of order $o(|| \hat{\bm{p}}_n-  \bm{p}||^2)$. Thus, this shows an interesting connection between classical independence tests and the MI.
In general, by using the $\delta$-method, we have shown that the MI and other divergence measures can be expressed as a sum of a normal random variables plus a linear combination of chi-squares, neglecting the remainder that tends to 0. Also, this property may be used  to compute confidence intervals for a given divergence measure. Note that the normal term and the linear combination of chi-squares term are not independent since both depended on $\lVert \hat{\bm{p}}_n- \bm{p} \lVert$. In that case, one can use a first order approximation or compute explicitly the variance of the sum of the two terms, by applying the variance operator to the Taylor expansion.
%
In the independent case the normal term vanishes remaining the linear combination of chi-squares.

Finally we have briefly explored the MI geometry and proposed other information based measures, which may improve the power of the independence test. Using symbolic software and the $\delta$-method many interesting options can be explored. In addition, symbolic software may allow the use of higher order derivatives. This requires further research.

\section*{Acknowledgements}
The first author would like to acknowledge vice dean Maria Miroiu for her warm and hospitable welcome at the Department of Mathematics and Computer Science of Politehnica Bucharest, where this research was conducted.

\begin{appendix}

\newpage
\section{Appendix}\label{appendix}

\subsection{Calculation of the MI gradient}\label{ap:grad}

We first compute the MI partial derivatives without imposing the restriction $\sum_{i,j=1}^{I,J} p_{ij}=1$, and after we compute the partial derivatives with the restriction by using the chain rule. 

The partial derivative respective to a generic $p_{ij}$ is:
%
\begin{align*}
    \frac{\partial \widetilde{\text{MI}}}{\partial p_{ij}} = \frac{\partial}{\partial p_{ij}} \Big(\sum_{i,j=1}^{I,J} p_{ij}\ln \frac{p_{ij}}{p_{i*}p_{*j}} \Big) &= 
    \ln \left(\frac{p_{i j}}{p_{i*}p_{*j}}\right)+1-\frac{p_{i j}}{p_i}-\frac{p_{i j}}{p_{i j}}-\sum_{\substack{k=1 \\ k \neq j}}^J \frac{p_{ik}}{p_{i *}}-\sum_{\substack{l=1 \\ l \neq i}}^I \frac{p_{lj}}{p_{* j}} & \\
    &=\ln \left(\frac{p_{i j}}{p_{i*}p_{*j}}\right)-\sum_{k=1}^J \frac{p_{ik}}{p_{i*}} - \sum_{l=1}^I \frac{p_{l j}}{p_{*j}}+1 = \ln \left(\frac{p_{i j}}{p_{i*}p_{*j}}\right) - 1
\end{align*}

where the tilde stands for the non-restricted MI. 

Let's denote the result $\ell_{ij}-1$. Now consider the parametrization $\vec{c}(p_{11}, \ldots, p_{12}, \ldots, p_{(I-1)J})= (\bm{p}, 1 - \displaystyle\sum_{\substack{i,j=1 \\ i \neq I, j \neq J}}^{I,J} p_{ij} )$.  Then,
\begin{align}\label{eq. grad}
    D(\MI)=D(\widetilde{\text{MI}} \circ \vec{c})(\bm{p}) = D \widetilde{\text{MI}}\bigg|_{\vec{c}(\bm{p})} \cdot D(\vec{c})\bigg|_{\bm{p}} &= (\ell_{11}-1, ...,\ell_{IJ}-1) \begin{pmatrix}
1 & 0 & 0 & \cdots & 0 \\
0 & 1 & 0 & \cdots & 0 \\
0 & 0 & 1 & \cdots & 0 \\
\vdots & \vdots & \vdots & \ddots & \vdots \\
0 & 0 & 0 & \cdots & 1 \\
-1 & -1 & -1 & \cdots & -1
\end{pmatrix} \\
&=(\ell_{11}-\ell_{IJ}, ...,\ell_{(I-1)J}-\ell_{IJ}) 
\end{align}


\subsection{Calculation of the MI Hessian matrix}\label{ap:hess}

Again, we compute the second-order derivatives without the normalization restriction and then include it. With careful consideration we can see that
\begin{equation*}
     \widetilde{H}_{ij, st}:=A_{ij, st}=\begin{cases}
        \frac{1}{p_{ij}}-\frac{1}{p_{i*}} -\frac{1}{p_{*j}} & \ \ s = i,\ t=j \\
          -\frac{1}{p_{i*}} & \ \ s = i,\ t \neq j  \\
           -\frac{1}{p_{*j}}  & \ \ s \neq i,\ t=j  \\
          \end{cases}
\end{equation*}

where the tilde stands for the non-restricted Hessian.

Looking at the MI gradient expression in eq. \ref{eq. grad} notice that we can rewrite it as $D(\MI) = (D(\widetilde{\text{MI}}) \circ \vec{c})(\bm{p}) + (1,\ldots, 1) - \ell_{IJ}(1,\ldots,1)$. Using this relation the Hessian matrix results in:
\begin{align}
    D^2(\MI)=D\Big(D(\widetilde{\text{MI}}) \circ \vec{c})(\bm{p}) + (1,\ldots, 1) - \ell_{IJ}(1,\ldots,1) \Big) = \widetilde{H}_{1:(I-1)J \times 1:IJ} \bigg|_{\vec{c}(\bm{p})} \begin{pmatrix}
1 & 0 & 0 & \cdots & 0 \\
0 & 1 & 0 & \cdots & 0 \\
0 & 0 & 1 & \cdots & 0 \\
\vdots & \vdots & \vdots & \ddots & \vdots \\
0 & 0 & 0 & \cdots & 1 \\
-1 & -1 & -1 & \cdots & -1 
\end{pmatrix} \notag \\ 
- D\big(\ell_{IJ}(1,\ldots, 1)\big)
\end{align}

with, 
$$D(\ell_{IJ})=\frac{\partial}{\partial p_{st}} (\ell_{IJ})=
\begin{cases}
        \frac{1}{p_{IJ}}-\frac{1}{p_{I*}}-\frac{1}{p_{*J}} & \ \ s \neq I, \ t\neq J\\
         \frac{1}{p_{IJ}}-\frac{1}{p_{*J}} & \ \ s = I, \ t\neq J\\
          \frac{1}{p_{IJ}}-\frac{1}{p_{I*}} & \ \ s \neq I, \ t= J\\
\end{cases}$$

Joining all together we get eq. \ref{eq: hess}.
\end{appendix}

\newpage

\bibliographystyle{IEEEtran}
\bibliography{bib}

\begin{thebibliography}{10}
\providecommand{\url}[1]{#1}
\csname url@samestyle\endcsname
\providecommand{\newblock}{\relax}
\providecommand{\bibinfo}[2]{#2}
\providecommand{\BIBentrySTDinterwordspacing}{\spaceskip=0pt\relax}
\providecommand{\BIBentryALTinterwordstretchfactor}{4}
\providecommand{\BIBentryALTinterwordspacing}{\spaceskip=\fontdimen2\font plus
\BIBentryALTinterwordstretchfactor\fontdimen3\font minus
  \fontdimen4\font\relax}
\providecommand{\BIBforeignlanguage}[2]{{%
\expandafter\ifx\csname l@#1\endcsname\relax
\typeout{** WARNING: IEEEtran.bst: No hyphenation pattern has been}%
\typeout{** loaded for the language `#1'. Using the pattern for}%
\typeout{** the default language instead.}%
\else
\language=\csname l@#1\endcsname
\fi
#2}}
\providecommand{\BIBdecl}{\relax}
\BIBdecl

\bibitem{shannon1948}
C.~E. Shannon, ``A mathematical theory of communication,'' \emph{The Bell
  system technical journal}, vol.~27, no.~3, pp. 379--423, 1948.

\bibitem{Boltzmann1896}
L.~Boltzmann, \emph{Vorlesungen über Gastheorie}.\hskip 1em plus 0.5em minus
  0.4em\relax Verlag von Johann Ambrosius Barth, 1896.

\bibitem{kullback1951}
S.~Kullback and R.~A. Leibler, ``On information and sufficiency,'' \emph{The
  annals of mathematical statistics}, vol.~22, no.~1, pp. 79--86, 1951.

\bibitem{leandro2006}
L.~Pardo, \emph{Statistical Inference Based on Divergence Measures},
  1st~ed.\hskip 1em plus 0.5em minus 0.4em\relax Chapman and Hall/CRC, 2006.

\bibitem{menendez2006dna}
M.~L. Men{\'e}ndez, J.~A. Pardo, L.~Pardo, and K.~Zografos, ``On tests of
  independence based on minimum $\varphi$-divergence estimator with
  constraints: an application to modeling dna,'' \emph{Computational statistics
  \& data analysis}, vol.~51, no.~2, pp. 1100--1118, 2006.

\bibitem{zografos1993}
K.~ZOGRAFOS, ``Asymptotic properties of $\varphi$-divergence statistic and its
  applications in contingency tables,'' \emph{Int. J. Math. Stat. Sci.},
  vol.~2, pp. 5--21, 1993.

\bibitem{leandro1994maestro}
M.~Salicru, D.~Morales, M.~L. Menendez, and L.~Pardo, ``{On the Applications of
  Divergence Type Measures in Testing Statistical Hypotheses},'' \emph{Journal
  of Multivariate Analysis}, vol.~51, no.~2, pp. 372--391, November 1994.

\bibitem{hutter2001}
M.~Hutter, ``Distribution of mutual information,'' \emph{Advances in neural
  information processing systems}, vol.~14, 2001.

\bibitem{ihler2004nonparametric}
A.~T. Ihler, J.~W. Fisher, and A.~S. Willsky, ``Nonparametric hypothesis tests
  for statistical dependency,'' \emph{IEEE Transactions on Signal Processing},
  vol.~52, no.~8, pp. 2234--2249, 2004.

\bibitem{gabor2007}
G.~J. Sz{\'e}kely, M.~L. Rizzo, and N.~K. Bakirov, ``{Measuring and testing
  dependence by correlation of distances},'' \emph{The Annals of Statistics},
  vol.~35, no.~6, pp. 2769 -- 2794, 2007.

\bibitem{gabor2009}
G.~J. Sz{\'e}kely and M.~L. Rizzo, ``Brownian distance covariance,'' \emph{The
  Annals of Applied Statistics}, vol.~3, no.~4, pp. 1236--1265, 2009.

\bibitem{gabor2013energy}
G.~J. Szekely and M.~L. Rizzo, ``Energy statistics: A class of statistics based
  on distances,'' \emph{Journal of statistical planning and inference}, vol.
  143, no.~8, pp. 1249--1272, 2013.

\bibitem{lopez2013}
D.~Lopez-Paz, P.~Hennig, and B.~Sch{\"o}lkopf, ``The randomized dependence
  coefficient,'' \emph{Advances in neural information processing systems},
  vol.~26, 2013.

\bibitem{reshef2011}
D.~N. Reshef, Y.~A. Reshef, H.~K. Finucane, S.~R. Grossman, G.~McVean, P.~J.
  Turnbaugh, E.~S. Lander, M.~Mitzenmacher, and P.~C. Sabeti, ``Detecting novel
  associations in large data sets,'' \emph{Science}, vol. 334, no. 6062, pp.
  1518--1524, 2011.

\bibitem{simon2014comment}
N.~Simon and R.~Tibshirani, ``Comment on" detecting novel associations in large
  data sets" by {R}eshef et. al.'' \emph{arXiv preprint arXiv:1401.7645}, 2014.

\bibitem{gorfine2012comment}
M.~Gorfine, R.~Heller, and Y.~Heller, ``Comment on “detecting novel
  associations in large data sets” by {R}eshef et. al.'' \emph{Science},
  2012.

\bibitem{arthur2005}
A.~Gretton, O.~Bousquet, A.~Smola, and B.~Sch{\"o}lkopf, ``Measuring
  statistical dependence with hilbert-schmidt norms,'' in \emph{Algorithmic
  Learning Theory}, S.~Jain, H.~U. Simon, and E.~Tomita, Eds.\hskip 1em plus
  0.5em minus 0.4em\relax Berlin, Heidelberg: Springer Berlin Heidelberg, 2005,
  pp. 63--77.

\bibitem{reimherr2013}
M.~Reimherr and D.~L. Nicolae, ``On quantifying dependence: A framework for
  developing interpretable measures,'' \emph{The Annals of Applied Statistics},
  vol.~28, no.~1, pp. 116--130, 2013.

\bibitem{vaart2000}
A.~W. Van~der Vaart, \emph{Asymptotic statistics}.\hskip 1em plus 0.5em minus
  0.4em\relax Cambridge university press, 2000, vol.~3.

\bibitem{vaart1996}
A.~W. Van Der~Vaart and J.~A. Wellner, \emph{Weak convergence}.\hskip 1em plus
  0.5em minus 0.4em\relax Springer, 1996.

\bibitem{kullback1997}
S.~Kullback, \emph{Information theory and statistics}.\hskip 1em plus 0.5em
  minus 0.4em\relax Courier Corporation, 1997.

\bibitem{ver2012}
J.~M. Ver~Hoef, ``Who invented the delta method?'' \emph{The American
  Statistician}, vol.~66, no.~2, pp. 124--127, 2012.

\bibitem{oehlert1992}
G.~W. Oehlert, ``A note on the delta method,'' \emph{The American
  Statistician}, vol.~46, no.~1, pp. 27--29, 1992.

\bibitem{rao1973}
C.~R. Rao, \emph{Linear statistical inference and its applications}.\hskip 1em
  plus 0.5em minus 0.4em\relax Wiley New York, 1973, vol.~2.

\bibitem{agresti2012}
A.~Agresti, \emph{Categorical data analysis}, 2nd~ed.\hskip 1em plus 0.5em
  minus 0.4em\relax John Wiley \& Sons, 2012.

\bibitem{mathai1992}
A.~Mathai and S.~B. Provost, \emph{Quadratic Forms in Random Variables}.\hskip
  1em plus 0.5em minus 0.4em\relax Marcel Dekker, 1992.

\bibitem{seber2008matrix}
G.~A. Seber, \emph{A matrix handbook for statisticians}.\hskip 1em plus 0.5em
  minus 0.4em\relax John Wiley \& Sons, 2008.

\bibitem{moschopoulos1984}
P.~Moschopoulos and W.~Canada, ``The distribution function of a linear
  combination of chi-squares,'' \emph{Computers \& mathematics with
  applications}, vol.~10, no. 4-5, pp. 383--386, 1984.

\bibitem{pearson1900}
K.~Pearson, ``X. {O}n the criterion that a given system of deviations from the
  probable in the case of a correlated system of variables is such that it can
  be reasonably supposed to have arisen from random sampling,'' \emph{The
  London, Edinburgh, and Dublin Philosophical Magazine and Journal of Science},
  vol.~50, no. 302, pp. 157--175, 1900.

\bibitem{baird1983}
D.~Baird, ``The fisher/pearson chi-squared controversy: A turning point for
  inductive inference,'' \emph{The British Journal for the Philosophy of
  Science}, vol.~34, no.~2, pp. 105--118, 1983.

\bibitem{fisher1922}
R.~A. Fisher, ``On the interpretation of $\chi$ 2 from contingency tables, and
  the calculation of p,'' \emph{Journal of the royal statistical society},
  vol.~85, no.~1, pp. 87--94, 1922.

\bibitem{neyman1933}
J.~Neyman and E.~S. Pearson, ``Ix. {O}n the problem of the most efficient tests
  of statistical hypotheses,'' \emph{Philosophical Transactions of the Royal
  Society of London. Series A, Containing Papers of a Mathematical or Physical
  Character}, vol. 231, no. 694-706, pp. 289--337, 1933.

\bibitem{cressie1989}
N.~Cressie and T.~R. Read, ``Pearson's x2 and the loglikelihood ratio statistic
  g2: a comparative review,'' \emph{International Statistical Review/Revue
  Internationale de Statistique}, pp. 19--43, 1989.

\bibitem{thomas2006}
M.~C. Thomas and A.~T. Joy, \emph{Elements of information theory}.\hskip 1em
  plus 0.5em minus 0.4em\relax Wiley-Interscience, 2006.

\bibitem{rubia2024hist}
J.~M. De~La~Rubia, ``Rice university rule to determine the number of bins,''
  \emph{Open Journal of Statistics}, vol.~14, no.~1, pp. 119--149, 2024.

\bibitem{mchugh2013chi}
M.~L. McHugh, ``The chi-square test of independence,'' \emph{Biochemia medica},
  vol.~23, no.~2, pp. 143--149, 2013.

\bibitem{studholme1999overlap}
C.~Studholme, D.~L. Hill, and D.~J. Hawkes, ``An overlap invariant entropy
  measure of 3d medical image alignment,'' \emph{Pattern recognition}, vol.~32,
  no.~1, pp. 71--86, 1999.

\bibitem{marius2011}
M.~Vila, A.~Bardera, M.~Feixas, and M.~Sbert, ``Tsallis mutual information for
  document classification,'' \emph{Entropy}, vol.~13, no.~9, pp. 1694--1707,
  2011.

\bibitem{tsallis1988}
C.~Tsallis, ``Possible generalization of boltzmann-gibbs statistics,''
  \emph{Journal of statistical physics}, vol.~52, pp. 479--487, 1988.

\end{thebibliography}
\end{document}